\documentclass[journal]{IEEEtran}

\usepackage[utf8]{inputenc}
\usepackage{amsmath, amsthm, amsfonts}
\usepackage{graphicx}
\usepackage{amssymb}
\usepackage{xcolor}
\usepackage{comment}
\usepackage{lipsum}
\usepackage{float}
\usepackage{bbold}
\usepackage{orcidlink}
\usepackage{booktabs}
\usepackage{multirow}

\DeclareMathOperator{\minimise}{minimize}
\DeclareMathOperator{\st}{subject \; to}

\begin{document}

%\title{Online Feedback Optimization with Regression-based Small-signal Stability Constraints}
\title{Small-Signal Stability Oriented Real-Time Operation of Power Systems with a High Penetration of Inverter-Based Resources}
%\title{Real-Time Small-Signal Stability in Power Systems: Overcoming the challenges of High Converter Penetration}

\author{Francesca~Rossi\orcidlink{0000-0002-4791-151X}, %~\IEEEmembership{Student~Member,~IEEE,}
        Juan~Carlos~Olives-Camps\orcidlink{0000-0002-3122-1223},~\IEEEmembership{Member,~IEEE,}
        Eduardo~Prieto-Araujo\orcidlink{0000-0003-4349-5923} ,~\IEEEmembership{Senior~Member,~IEEE}
        and~Oriol~Gomis-Bellmunt\orcidlink{0000-0002-9507-8278},~\IEEEmembership{Fellow,~IEEE,}
\thanks{Manuscript received Month day, year; revised Month day, year and Month day, year; accepted Month day, year.}%
\thanks{Grant TED2021-130351B-C22 (HP2C-DT) funded by MICIU/AEI/10.13039/501100011033 and by European Union NextGenerationEU/PRTR.}%
\thanks{All of the authors are with the Centre d’Innovació Tecnològica en Convertidors Estàtics I Accionaments research group (CITCEA-UPC), Universitat Politècnica de Catalunya, Barcelona 08034, Spain (e-mails: francesca.rossi@upc.edu, juan.carlos.olives@upc.edu, eduardo.prieto-araujo@upc.edu, oriol.gomis@upc.edu).}}

% The paper headers
%\markboth{IEEE TRANSACTIONS ON POWER SYSTEMS,~Vol.~X, No.~Y, MONTH~YEAR}%
%{ROSSI \MakeLowercase{\textit{et al.}}: title}

\maketitle

\begin{abstract}
This study proposes a control strategy to ensure the safe operation of modern power systems with high penetration of inverter-based resources (IBRs) within an optimal operation framework. The objective is to obtain operating points that satisfy the optimality conditions of a predefined problem while guaranteeing small-signal stability. The methodology consists of two stages. First, an offline analysis of a set of operating points is performed to derive a data-driven regression-based expression that captures a damping-based stability index as a function of the operating conditions. Second, an Online Feedback Optimization (OFO) controller is employed to drive the system toward an optimal operating point while maintaining a secure distance from the instability region. The proposed strategy is evaluated on an academic test case based on a modified version of the IEEE 9-bus system, in which synchronous generators are replaced by IBRs operating under both grid-following and grid-forming control modes. The results demonstrate the effectiveness of the method and are discussed in detail.
\end{abstract}


\IEEEpeerreviewmaketitle

\section{Introduction} \label{sec:intro}
\subsection{Context and Motivation}
The increasing penetration of renewable energy sources (RES), combined with the progressive phase-out of fossil-fuel-based synchronous generation, is transforming the electrical grid into a power-electronics-dominated system. This transition is profoundly changing the grid’s dynamic behavior: traditional synchronous generators (SGs) inherently provide inertia and damping to electromechanical oscillations, whereas inverter-based resources (IBRs) exhibit fast dynamics governed by their internal control loops. As a result, modern power systems exhibit reduced inertia and are dominated by faster, control-driven phenomena occurring on the timescale of converter dynamics~\cite{StabilityTop,milano2018foundations}. %10$\micro$s-10 ms.
Moreover, for both environmental and economic reasons, system operation is incentivized to maximize the utilization of renewable generation, particularly from non-dispatchable sources such as photovoltaic (PV) and wind power. This trend further pushes power systems to operate closer to their stability limits, where small disturbances or unforeseen fluctuations in generation and demand can lead to dynamic instabilities. Recent European network blackout events, such as those that occurred in the Iberian Peninsula and North Macedonia, serve as illustrative examples. Although investigations into the root causes of these events are ongoing, early analyses suggest that they were not caused by physical damage to infrastructure or extreme weather conditions, but rather occurred under normal operating conditions — one during periods of high renewable generation~\cite{entsoe2025_iberian_factual}, and another during low nighttime weekend demand~\cite{entsoe2025_northMacedonia_incident}. These events have highlighted that transmission system operators (TSOs) currently lack adequate real-time tools to monitor and manage stability margins under rapidly changing operating conditions.

To address these challenges, numerous studies have proposed incorporating stability constraints, both static and dynamic, into the optimal power flow (OPF) problem. Following the conventional classification of power system stability~\cite{StabilityTop}, different stability-constrained OPF formulations can be defined. Transient, voltage, and frequency stability-constrained OPF addresses the effects of large disturbances, providing the optimal pre-contingency generators dispatch able to withstand contingencies as faults and outages~\cite{book_scopt}. When considering small perturbations, small-signal stability-constrained OPF (SSSC-OPF) can be employed. However, if the small-signal model is properly formulated to capture electromagnetic transient (EMT) phenomena, SSSC-OPF is not limited to prevent instabilities due to small-disturbances, but can also detect and prevent from converter-driven and resonance-induced instabilities. Consequently, the SSSC-OPF formulation ensures that the system does not exhibit undamped oscillations during normal operation or under small disturbances.

Concerning the real-time implementation of optimal operation algorithms, Online Feedback Optimization (OFO) has emerged as a promising approach. The fundamental concept of OFO is the implementation of optimization algorithms as feedback controllers, which are coupled to physical plants via a closed-loop scheme \cite{Florian2020}. Through this interconnection, the controller drives the operating point towards the solution of the optimization problem using real-time measurements instead of computing the equations that describe the system's behavior. As a result, the system exhibits robustness to disturbances and uncertainties. Furthermore, by circumventing the numerical evaluation of the nonlinear plant model, the computational burden is significantly reduced while preserving the industrial property \cite{olives2025uac}.

To contextualize the proposed approach, the following section reviews relevant literature on SSSC-OPF formulations and the application of OFO in power system operation.

\subsection{Literature Review}
\subsubsection{Small-signal Stability-Constrained Optimal Power Flow}
%In this context, small-signal stability-constrained optimal power flow (SSSC-OPF) has attracted increasing attention in recent years, and 
Numerous formulations have been proposed in the literature to improve the tractability of the SSSC-OPF problem by circumventing the explicit inclusion of the full set of differential algebraic equations (DAE) describing system dynamics~\cite{stabilityconstrained,Milano2011,thams2017data}. Some approaches rely on the linearized state-space representation of these DAE and the corresponding eigenvalue analysis, as commonly employed in small-signal stability assessment~\cite{kundur}. However, the relationship between the eigenvalue-based indicators (e.g., damping ratios and real parts of eigenvalues) and the system states and parameters is implicit~\cite{SSSCOPF_SVM_2022}.
To address this issue, several studies have proposed sensitivity-based formulations, where the stability constraint is expressed in terms of the linear sensitivities of the system state matrix eigenvalues with respect to small variations in generators' active power~\cite{Milano2011,7539295}. Despite limiting the analysis to a number of critical eigenvalues, the recalculation of the sensitivities at each iteration remains a necessity, making these approaches unsuitable for real-time operation.
%Although both~\cite{Milano2011,7539295} suggest reducing computational effort by restricting the sensitivity analysis to a limited number of critical eigenvalues, they still require recalculating the sensitivities at each iteration, making them unsuitable for real-time operation.

The employment of machine learning (ML)-based surrogate models can facilitate the formulation of stability constraints, thereby alleviating the computational burden~\cite{hasan2020survey}.
%To alleviate the computational burden, machine learning (ML)-based surrogate models can be employed to formulate the stability constraint~\cite{hasan2020survey}. 
This approach, generally named Constraint Learning (CL)~\cite{maragno2023mixed,fajemisin2024optimization}, is particularly effective when the analytical expression of the intended constraint is unknown or complex~\cite{fajemisin2024optimization}. Moreover, ML techniques shift the most computationally demanding tasks to the offline phase, during which the model is trained and the necessary training data are generated and collected. Once trained, the model can be used online to provide a fast and efficient evaluation of the stability constraint. As reviewed in~\cite{maragno2023mixed}, commonly used ML models for CL include both linear approaches, such as linear regression and Support Vector Machines (SVMs) with linear kernels, and nonlinear models, including Decision Trees (DTs), ensemble methods, and Neural Networks (NNs). While these nonlinear models can capture the behavior of small-signal stability indices, their integration into OPF formulations typically requires linearization techniques involving binary variables~\cite{maragno2023mixed}, as demonstrated in~\cite{9141308,thams2017data}. This results in mixed-integer nonlinear programming (MINLP) problems, whose solution for large-scale power systems remains computationally challenging and often intractable~\cite{capitanescu2011state}. In~\cite{liu2021explicit}, an SVM is employed to formulate a linear small-signal stability constraint with good computational efficiency. However, the method relies on locally trained models built around each unstable operating point, making it suitable for corrective re-dispatch rather than real-time preventive control. \\
In a previous work by the authors~\cite{rossiSSSCOPF2024}, the performance of linear models was compared against Multivariate Adaptive Regression Splines (MARS)~\cite{friedman1991multivariate}, a nonlinear regression technique identified as suitable for inclusion in OPF formulations. MARS provides a continuous and differentiable analytical expression (class $\mathcal{C}^1$)~\cite{friedman1991multivariate}, making it compatible with OPF problems solved via sequential quadratic programming (SQP), in contrast to the previously mentioned nonlinear models. Moreover, MARS demonstrated superior robustness, with respect to the linear models, in estimating small-signal stability indicators across OPF formulations with different objective functions. %For these reasons, MARS is also adopted in this work.
\subsubsection{Online Feedback Optimization in Power Systems Operation} 
Up to now, OFO has been mainly investigated in the context of distribution system operation, where its advantages are particularly significant due to the lack of accurate grid models and the uncertainty associated with RES. Applications in the literature include voltage regulation~\cite{ORTMANN2020106782,olives2022ac} and optimal reactive power dispatch~\cite{6930779}. More recently, its applicability to transmission systems has also been demonstrated, where its main advantages lie in the smooth, continuous control actions it enables against disturbances. In particular, feedback-based online optimization has been shown to provide stable tracking of optimal operating points in transmission networks, effectively combining frequency regulation and economic dispatch in a single real-time control loop~\cite{colombino2019online,menta2018stability}. Large-scale studies on real subtransmission grids further show that OFO can enforce voltage and line-flow limits, handle discrete devices such as tap changers, and reduce renewable curtailment even under model uncertainty~\cite{ortmann2026subtransmission}. Although TSOs typically possess system models, their accuracy may be limited, and relying directly on real-time measurements can offer a more robust and adaptive alternative~\cite{ofo_tso_dso_2025}. Furthermore, incorporating small-signal stability constraints into OFO can also be relevant for distribution systems operating in islanded or weakly connected modes, provided that suitable small-signal models (obtained through dynamic system identification or black-box modeling) are available.

\subsection{Contributions}
This paper presents a novel framework that integrates OFO with ML-based small-signal stability constraints to enable real-time, stability-constrained power system operation. The main contributions are summarized as follows:
\begin{itemize}
\item Development of a Small-Signal Stability-Constrained Online Feedback Optimization (SSSC-OFO) formulation capable of real-time generator dispatch adjustment to ensure static and dynamic stability.
\item Comparative evaluation of the proposed SSSC-OFO against a conventional SSSC-OPF in terms of performance and computational efficiency.
\item Application and validation on a power-electronics-dominated test system comprising grid-following (GFL) and grid-forming (GFM) converters with minimized synchronous generation.
\end{itemize}

\subsection{Paper Organization}
The paper is organized as follows. Section~\ref{sec:method} describes the three main components of the proposed methodology: the formulation of the small-signal stability constraint using data-driven regression, the formulation of the OFO framework, and their integration into the proposed SSSC-OFO scheme. Section~\ref{sec:results} presents the application of the proposed method to a test system and discusses the obtained results in detail. Finally, Section~\ref{sec:concl} provides concluding remarks and outlines directions for future research.

\section{Methodology Description}\label{sec:method}

\subsection{Small-Signal Stability Constraint Formulation}
\label{ssec:MARS}
To formulate a constraint on system dynamics for inclusion in the optimal operation problem, a regression-based approach is employed. The use of regression provides a compact surrogate model  of an otherwise complex problem. %Small-signal stability analysis can be performed using linearized state-space models, as commonly applied in eigenvalue-based stability assessment. However, such a formulation becomes unwieldy for large power systems, as it requires handling a large number of equations [cite spyros??]. In contrast, 
The regression-based surrogate is trained to provide the same outcome as the conventional eigenvalue-based analysis of the linearized state-space of the system, while offering a significantly simpler formulation that reduces the dimensionality of the problem. Moreover, regression techniques provide flexibility in selecting the mathematical representation most suitable depending on how the optimization problem is formulated. %In the authors’ previous work on small-signal stability-constrained optimal power flow (OPF)~\cite{rossiSSSCOPF2024}, which focused on implementing a formulation using an SQP optimization method, specific requirements for the regression model were identified: the need for an analytical expression that is continuous and differentiable. These criteria pertain to linear and polynomial models. Among non-linear alternatives, Multivariate Adaptive Regression Splines (MARS) [cite friedmann] were identified as a particularly effective technique. The study further demonstrated that MARS achieves superior and more robust accuracy, particularly under variations of the OPF objective function.
In this work, MARS is employed for its ability and robustness in capturing the nonlinear relation between the OPF control variables and the stability indicator~\cite{rossiSSSCOPF2024}, and for providing an analytic expression of class $\mathcal{C}^1$, suitable for formulating a smooth constraint in an SQP-based optimization.
%In the present study, focused on small-signal stability-constrained optimal system operation via OFO, the nature of the stability assessment remains the same, and the requirement for an analytical, differentiable formulation still applies, in order to solve the OFO through SQP. Therefore, MARS is adopted in this work as the regression-based model for embedding small-signal stability constraints.

To develop the regression model, a representative training dataset is required. This dataset must capture the mapping between system operating conditions and small-signal stability across the entire operable space, while providing finer granularity near the stability boundary — i.e., the region where both stable and unstable operating points occur. A dataset with such features needs to be generated synthetically~\cite{venzke2021efficient}, by performing stability analyses for a large number of operating points sampled within the operable space. Formally, the training dataset can be expressed as $\mathcal{D} = (\mathbf{X} \mid \mathbf{z})$. The matrix $\mathbf{X} $ contains the system state variables for each sampled operating point, including bus voltage magnitudes and phases, as well as generators and loads active and reactive power injections. The vector $\mathbf{z}$ stores the corresponding stability assessment outcomes. Since regression requires a continuous numerical target, $\mathbf{z}$ is expressed in terms of a stability index. For small-signal stability, this can be the Damping Index (DI) defined in~\cite{collados2022stability}, given as
\begin{equation}
    \text{DI} = 1 -\min\{\xi_1,...,\xi_{|\lambda|}\} \label{eq:DI},
\end{equation}
\noindent where $\xi_i$ denotes the damping ratio of the $i$-th complex eigenvalue $\lambda_i$, and $|\mathbf{\lambda}|$ is the number of critical eigenvalues, as defined in~\cite{rossiML2022}. Given this formulation of the DI, small-signal stability is satisfied if $\text{DI} < 1$. Moreover, a stricter requirement can be imposed by introducing a threshold $\theta < 1$, such that $\text{DI} < \theta$, thereby enforcing a prescribed damping margin.

Once the training dataset is generated, the MARS model is fitted. MARS expresses the regression function as a linear combination of piecewise-linear basis functions, known as hinge functions. For the $i$-th variable $\mathbf{x}_i \in \mathbf{X} $, a hinge function is defined as
\begin{equation}
h(\mathbf{X}) = \max\{0, \mathbf{x}_i - t\} \quad \text{or} \quad \max\{0, t - \mathbf{x}_i\}, \label{eq:hinge}
\end{equation}
where $t$ is a point within the range of $\mathbf{x}_i$ in the training dataset, referred to as a knot. Although each basis function depends only on the single variable $\mathbf{x}_i$, it is formally regarded as a function defined over the entire input space~\cite{hastie2009elements}. During training, the MARS algorithm automatically determines the placement of knots and selects the hinge functions that contribute most to improving regression accuracy while ensuring generalization. The resulting regression model takes the form
\begin{equation}
g(\textbf{X}) = \beta_0 + \sum_{m=1}^M \beta_m h_m(\textbf{X}),
\end{equation}
where $\beta_0$ is the intercept, $\beta_m$ are the coefficients of the hinge functions $h_m(X)$, and $M$ is the number of selected basis functions. 
Finally, within an optimization problem, the guarantee of small-signal stability is provided by including the constraint
\begin{equation}
    g(\textbf{X})<\theta \quad \text{with}\quad \theta\leq 1 \label{eq:stab_constr}.
\end{equation}

%\cite{rossiML2022} % Data Generation Methodology for Machine Learning-based Power System Stability Studies

%\cite{rossiSSSCOPF2024} % Optimal Power Flow With Regression-Based Small-Signal Stability Constraints 

\subsection{Online Feedback Optimization} \label{ssec:OFO}

The OFO technique is based on the integration of the dynamics of a numerical optimization solver and the system dynamics to be controlled within a closed-loop structure. In this configuration, the solver is employed to update the control inputs for the system in response to the measurements obtained from the system. Note that this approach does not require an accurate plant model since is based on real-time measurements. Instead, it only requires the system's input-output sensitivities, which can be obtained on site \cite{olives2022ac}. Therefore, the performance of the controller remains unaffected by model inaccuracies and is robust to disturbances and uncertainty, as these effects are implicit in the measurements.

The objective of OFO-based techniques is to steer the power system towards a steady-state optimal operating point. Therefore, under certain assumptions concerning the separation of time scales and the rapid mitigation of transients \cite{OFO_stability2018}, the power system may be solely defined by its steady-state behavior. In this context, consider a set of nonlinear algebraic equations $\mathbf{f}(\cdot)$, as follows:
\begin{equation} \label{eq:steadystate}
    \mathbf{y} = \mathbf{f}( \mathbf{u,w} ),
\end{equation}
where the vector $\mathbf{y} \in \mathbb{R}^n$ represents the system outputs or measurements, while the vectors $\mathbf{u} \in \mathbb{R}^p$ and $\mathbf{w} \in \mathbb{R}^q$, respectively, represent the controlled and exogenous system inputs. The set of equations  \eqref{eq:steadystate} includes, but is not limited to, the equations that describe the power flow problem and the stationary effect of the controllers implemented in the IBRs and SGs. It should be noted that the specific expressions of \eqref{eq:steadystate} are irrelevant to the OFO-based controller design process, and they are only included for clarification purposes.

To obtain the OFO-based control law that optimizes the operating point of the system, we first need to define the following optimization problem:
\begin{subequations}
\begin{align}
    \underset{\mathbf{u}}{\minimise} \hspace{2mm} & \gamma \phi(\mathbf{y}, \mathbf{u}),\tag{5a}\label{eq:objfun}\\
    \st \quad
&\mathbf{y} - \mathbf{f}(\mathbf{u,w}) = \mathbf{0}, \tag{5b}\label{eq:eqconstr}\\
%& g(\mathbf{y,w}) <\theta \tag{5c} \label{eq:ineqnlinconstr} \\
& \mathbf{u} \in \mathcal{U}, \tag{5c} \label{eq:ineqlinconstrU}\\
& \mathbf{y} \in \mathcal{Y},\tag{5d} \label{eq:ineqlinconstrY}
%\label{eq:genform}
\end{align}
\end{subequations}

% \begin{equation} \label{eq:genform}
%     \begin{split}
%         \underset{\mathbf{u}}{\minimise} \hspace{2mm} & f(\mathbf{y}, \mathbf{u}), \\
%         \st \hspace{2mm} & \mathbf{y} - \mathbf{h}(\mathbf{u,w}) = \mathbf{0}, \\
%         & g(\mathbf{u,y,w}) <\theta, \\
%         & \mathbf{u} \in \mathcal{U}, \\
%         & \mathbf{y} \in \mathcal{Y},
%     \end{split}
% \end{equation}

\noindent where $\phi(\cdot)$ is the scalar objective function, $\gamma$ is a scaling factor enhancing the convexity of the objective function, the nonempty sets $\mathcal{Y}$ and $\mathcal{U}$ represent the feasible hyper-spaces of the corresponding variables. %are the damping factor and the corresponding equation obtained by the MARS method, as detailed in Section \ref{ssec:MARS}.
%The objective of this work is to minimise the contribution of coal-based generators in order to reduce CO$_2$ emissions into the atmosphere. The result of this is an increased participation of IBRs in the system, which is known to lead to unstable operating points. Therefore, to guarantee the stability of the system, it is necessary to add a second term to the objective function. This term negatively weights the damping coefficient that characterises the system ($\theta$).
The constraint in \eqref{eq:eqconstr} stipulates that the operating point must be one that is physically feasible for the system. Whereas, %the constraint in \eqref{eq:ineqnlinconstr} establishes how to determine the damping factor that characterises the stability. Finally, 
the purpose of the two constraints in \eqref{eq:ineqlinconstrU}-\eqref{eq:ineqlinconstrY} is to ensure that the variables remain within a feasible space from the engineering point of view. These constraints are typically expressed as box-constraints. Thus, both sets of constraints can be expressed linearly as follows:
\setcounter{equation}{5}
\begin{equation} \label{eq:ineqlincon}
    \begin{split}
        & \mathbf{b} - \mathbf{A u} \geq \mathbf{0}, \\
        & \mathbf{d} - \mathbf{C y} \geq \mathbf{0},
    \end{split}
\end{equation}
where the vectors $\mathbf{b}$ and $\mathbf{d}$ encapsulate the limits imposed on the variables, whereas matrices $\mathbf{A}$ and $\mathbf{C}$ contain the constants that scale and change the sign of the variables.

The nonconvexity of the optimization problem \eqref{eq:objfun}-\eqref{eq:ineqlinconstrY} is addressed through the implementation of a projected gradient flow algorithm as presented in \cite{haberle2020non}. Starting from the control action $\mathbf{u}$ in the current iteration, the control algorithm proceeds to update the value ($\mathbf{u}^+$) using a projected gradient step method with a fixed step size\footnote{A fixed step size algorithm must be used as an adverse consequence of not evaluating the set of equations $\mathbf{f}(\cdot)$ numerically.} as follows:
\begin{equation}\label{eq:u_update}
    \mathbf{u}^+ = \mathbf{u} + \alpha \hat{\boldsymbol{\sigma}}(\mathbf{u,y}),
\end{equation}
where $\alpha$ is a nonnegative fixed step-size, and $\hat{\boldsymbol{\sigma}}(\cdot)$ is the direction vector calculated as follows:
\begin{subequations} \label{eq:sigma}
\begin{align}
    \hat{\boldsymbol{\sigma}}(\mathbf{u,y}) := \arg \min_{\boldsymbol{\delta} \in\mathbb{R}^p} & \; \| \boldsymbol{\delta} + \gamma \mathbf{G}^{-1}\mathbf{F}^{\top} \nabla \phi(\mathbf{y,u})^{\top} \|_\mathbf{G}^2 \tag{8a} \label{eq:sigma_1}\\
    \st & \; \mathbf{b} - \mathbf{A}(\mathbf{u} + \alpha \boldsymbol{\delta}) \geq \mathbf{0}, \tag{8b}\label{eq:sigma_2}\\
    & \; \mathbf{d} - \mathbf{C}(\mathbf{y} + \alpha \nabla_u \mathbf{f} \boldsymbol{\delta}) \geq \mathbf{0}, \tag{8c}\label{eq:sigma_3}
\end{align} 
\end{subequations}
where $\boldsymbol{\delta}$ is an auxiliary vector that corresponds to the projection of the optimal direction onto the feasible hyperspace, $\mathbf{F}^\top = \left[ \mathbb{I}_p \, \nabla \mathbf{f}^\top \right]$ corresponds to the sensitivity matrix, which appears as a consequence of the chain rule, and $\mathbf{G}$ weights the changes in the control variables.

Equations \eqref{eq:u_update} and \eqref{eq:sigma} define the structure of the OFO-based controller. As mentioned above, note that the computation of $\mathbf{f}(\cdot)$ does not appear in the controller structure; only the sensitivity matrix value, $\mathbf{F}$, is required. It is also noteworthy that the solution to problem \eqref{eq:sigma}, obtained at each iteration, guarantees that the point to which the system moves satisfies the constraint sets \eqref{eq:ineqlincon}, under the consideration that the increment is sufficiently small and can be approximated linearly.

Finally, note that due to its integral structure, when the controller reaches a steady state ($\mathbf{u}^+ = \mathbf{u}$), the value of $\hat{\boldsymbol{\sigma}}(\cdot)$ must be null (considering that $\alpha$ is a constant and positive value). For this to occur, the value of $\nabla_u \phi(\mathbf{y,u})$ must be equal to zero, indicating that the solution is a critical point of $\phi(\cdot)$ or there exist no projection onto the feasible region, which means that the point resides at the boundary of this region and the direction of descent is perpendicular to the tangent of the boundary at that point.

\subsection{Small-signal Stability Constrained Online Feedback Optimization}
The OFO formulation presented in the previous section does not guarantee small-signal stability of its solutions, since no constraints on system dynamics are incorporated. To ensure stability, the regression-based constraint in \eqref{eq:stab_constr} must be added to the optimization problem defined in \eqref{eq:objfun}–\eqref{eq:ineqlinconstrY}. 
% To reduce the computational burden of the problem, a penalty method is employed. This approach eliminates the explicit nonlinear inequality constraint %\eqref{eq:stab_constr} 
% by incorporating a penalty term into the objective function~\cite{citepenalty}. Consequently, the optimization problem can be reformulated as
% \begin{subequations}
% \begin{align}
%     \underset{\mathbf{u}}{\minimise} \quad 
%     & f'(\mathbf{y},\mathbf{u}) 
%       = \gamma \cdot (f(\mathbf{y}, \mathbf{u}) + \nonumber\\
%     &\quad + \beta\cdot\left[\max\{0, g(\mathbf{u}, \mathbf{y}, \mathbf{w}) - \theta\}\right]^n),
%     \tag{9}\label{eq:objfun_pen} \\
%     \st \quad 
%     & \eqref{eq:eqconstr}, \eqref{eq:ineqlinconstrU}, \eqref{eq:ineqlinconstrY} \nonumber
% \end{align}
% \end{subequations}

\begin{subequations}
\begin{align}
    \underset{\mathbf{u}}{\minimise} \quad 
    & \gamma \cdot \phi(\mathbf{y}, \mathbf{u}) 
    \nonumber \\
    \st \quad 
    & \eqref{eq:eqconstr}, \eqref{eq:ineqlinconstrU}, \eqref{eq:ineqlinconstrY} \nonumber\\
    & \theta - g(\mathbf{y},\mathbf{u},\mathbf{w})>0 \label{eq: ss_consr}
\end{align}
\end{subequations}
\noindent where $g(\cdot)$ is expressed as function of the variables $\{\mathbf{u,y,w}\}\subset \mathbf{X}$. %, $\beta$ is a weighting factor that penalizes dynamic instability, and $\gamma$ is a scaling factor introduced to enhance the convexity of the objective function. This reformulated objective function does not substantially alter the formulation in \eqref{eq:sigma_1}-\eqref{eq:sigma_3}; only in \eqref{eq:sigma_1} $\nabla f(\mathbf{y},\mathbf{u}) $ needs to be replaced by $\nabla f'(\mathbf{y},\mathbf{u},\mathbf{w})$. The presence of $\mathbf{w}$ in $f'$ does not affect the formulation, since the gradients are taken with respect to $\mathbf{y}$ and $\mathbf{u}$. Naturally, the gradient of the penalty term must also be computed. This requires evaluating both the gradient of the penalty term itself and the gradients of the hinge functions contained in $g(\cdot)$. Since the penalty term follows a hinge-like structure, its treatment is analogous to that of hinge functions. Accordingly, in \eqref{eq:hinge_deriv} the method for calculating the partial derivative of hinge-type expressions is illustrated, namely
% \begin{equation}
%     \frac{d h(x_i,t)}{dx_i} = \mathrm{max}\{0,\mathrm{sign}(x_i-t)\} \label{eq:hinge_deriv}
% \end{equation}. 

\section{Numerical evaluation} \label{sec:results}%and discussion

%The second scenario, designated the "industrial case," represents the IEEE 118-bus benchmark network model. The objective of the tests conducted on this network is to evaluate the scalability and performance of the presented algorithm.

This section presents the application of the proposed methodology to an academic test system, namely a modified version of the IEEE 9 bus system. This case study enables a clear illustration and visualization of the main insights underlying the methodology. Furthermore, the validation of the proposed SSSC-OPF formulation is provided.

\subsection{Test System}
The simulation test bench is based on a modified version of the IEEE 9-bus system~\cite{anderson2003power}. It represents a three-phase high-voltage transmission network consisting of nine buses, six transmission lines, three transformers, three load substations, and three generators. In contrast to the original system, which includes three SGs, two SGs are replaced with IBRs to emulate a scenario with high penetration of power electronics devices. Furthermore, to account for the impact of control strategies on system stability, one IBR is operated under GFM control while the other is operated under GFL control. 

\begin{figure}
    \centering
    \includegraphics[width=0.9\linewidth]{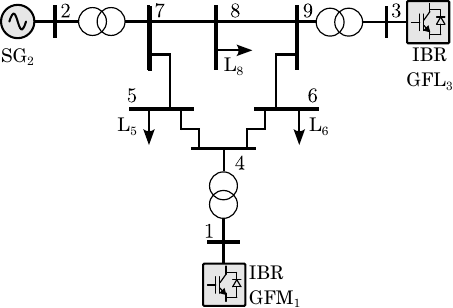}
    \caption{One-line diagram of the 9-bus system.}
    \label{fig:9bus}
\end{figure}

%The power system comprises eight substations. These are supplied by one VSC-based HVDC link and two conventional thermal power plants. 
Figure \ref{fig:9bus} shows the one-line diagram of the system under analysis. %, whereas Tables \ref{tab:gen_parameters}-\ref{tab:load_params} provide a summary of the main parameters employed to model it mathematically. 
The electrical lines of the system are considered symmetrical and are modeled using a $\pi$-model. The character of the lines is predominantly inductive, and all parameters are obtained from~\cite{anderson2003power}. Power substations are modeled as constant impedance balanced loads (R-L shunts load model). Table~\ref{tab:load_params} summarizes the main characteristics of the substations demand, including the load participation factors (i.e., the percentage of the total system demand allocated to each load), the load power factors, and the overall range of system demand.
%The VSC-based HVDC link is modelled using one single IBR, and each power plant is represented by a traditional synchronous generator.
Concerning the generators, it is assumed that both SG and IBRs are equipped with a primary power source that allows them to operate effectively at any feasible operating point. In addition, all generators have a communication system with a centralized control layer that calculates their setpoints. At the local level, the SG is incorporated with a governor and an AVR to regulate its steady-state outputs to the values dispatched by the proposed control layer. Therefore, the setpoints for SGs are defined as the injected active power and the generated voltage amplitude, measured at the bus to which they are connected. The IBR connected to bus 1 is configured to operate in a GFM mode, and therefore indicated as $\text{GFM}_1$. The device is equipped with an internal current control loop and an external voltage control loop. In addition, it contains an active power droop control stage, employed for synchronization, and reactive power droop. The IBR connected to bus 3 is configured to operate in a GFL mode, and therefore indicated as $\text{GFL}_3$. The GFL control scheme comprises a Phase-Locked Loop (PLL), an inner current control loop, an outer control loop for active power fed by a frequency droop control, and an outer reactive power control fed by a voltage droop. Table~\ref{tab:gen_parameters} summarizes the main generator parameters, including their rated power, rated voltage, and the assumed operating active power ranges. % for active and reactive power.

%This simplified system is employed to examine the impact of the design parameters on the methodologies that configure the controller and its performance.

\begin{table}[htbp]
\caption{Main Loads Parameters}
\label{tab:load_params}
\centering
\begin{tabular}{lccc}
\toprule
 & Min [MW]
 & Max [MW] & $\cos(\phi)$ \\
\midrule
Total Demand & 215.4 & 775.5 & 0.98 \\ \midrule
 & $\text{L}_5$ & $\text{L}_6$ & $\text{L}_8$ \\ \midrule
\begin{tabular}[c]{@{}l@{}}Load Participation Factor\\ {[\% Total Demand]}\end{tabular}
 & 40\% & 28\% & 32\% \\
\bottomrule
\end{tabular}
\end{table}

\begin{table}[hptb]
\caption{Main Generators Parameters}
\label{tab:gen_parameters}
\centering
\begin{tabular}{cccc}
\toprule
  & $\text{GFM}_1$ & $\text{SG}_2$ & $\text{GFL}_3$ \\
 \midrule
Rated Power [MVA] & 512 & 270 & 125 \\ \midrule
\begin{tabular}[c]{@{}c@{}}Active Power Operating Range \\{[\% Rated Power]}\end{tabular} & \multicolumn{3}{c}{20\% -- 95\%} \\ \midrule
%Active Power Operating Range {[\% Rated Power]} & \multicolumn{3}{c}{20\% -- 95\%} \\ \midrule
%\begin{tabular}[c]{@{}l@{}}Reactive Power\\ Operating Range\end{tabular} & \multicolumn{3}{c}{$\cos(\varphi) \leq 0.95$} \\ \midrule
%Reactive Power Operating Range & \multicolumn{3}{c}{$\cos(\varphi) \geq 0.95$} \\ \midrule
Rated RMS Voltage {[kV]} & 24 & 18 & 15.5 \\ \bottomrule
%\begin{tabular}[c]{@{}l@{}}Rated RMS \\ Voltage {[kV]}\end{tabular} & 24 & 18 & 15.5 \\ \bottomrule
\end{tabular}
\end{table}

\subsection{Optimization Problem}
The optimization problem formulated in this study aims to address two fundamental objectives of power system operation: efficiency and sustainability, while accounting for system limitations and both static and dynamic stability constraints. Accordingly, the objective function seeks to minimize power losses and reduce the active power output of SG$_2$, considering it as representing a fossil-fuel-based thermal power plant. The resulting objective function is therefore defined as a weighted sum of the squared power outputs of the generators, as in
\begin{equation}
    \phi(\mathbf{y},\mathbf{u}) = P_{\text{GFM}_1}^2 + 10\cdot P_{\text{SG}_2}^2 +P_{\text{GFL}_3}^2 \label{eq:obj_fun_cs}
\end{equation}
\noindent where the weighting factor associated with $P_{\text{SG}_2}$ penalizes fossil-based generation, and the use of squared power terms enhances the convexity of the objective function. Moreover, the values of the injected power are expressed in per unit with respect to generators nominal power, to fairly compare all generators regardless of its size.

The system limitations include generator capability boundaries (as defined by the active power operating range in Table~\ref{tab:gen_parameters}), whereas the static stability constraints enforce voltage magnitude limits (0.9 p.u. $\leq V\leq$ 1.1 p.u.).
The dynamic stability constraint is designed to ensure that the system maintains a minimum level of small-signal damping required for stable operation. Accordingly, $\theta = 1-\epsilon$ (with $\epsilon=1e-5$) is set in \eqref{eq: ss_consr}, ensuring a $\text{DI} < 1$. Following the proposed SSSC-OFO formulation, the setup described below is adopted.
\begin{itemize}
    \item Power flow configuration: Bus 1 is designated as the system slack bus, buses 2 and 3 are modeled as PV buses, and the remaining buses operate as PQ buses.
    \item Control actions: The controllable generator power setpoints and the voltage setpoints of all generators constitute the vector of control actions, defined as
    \begin{equation}
        \mathbf{u}=[P_{\text{SG}_2},P_{\text{GFL}_3},V_1,V_2,V_3].
    \end{equation}
    \item System outputs: %The remaining system quantities are treated as measurable outputs. 
    The output vector $\mathbf{y}$ includes both the variables subject to operational constraints and other measurable quantities of interest, defined as
\end{itemize}
\begin{equation}
    \mathbf{y} = [P_{\text{GFM}_1}, P_{\text{SG}_2}, V_1,V_2,V_3, V_4, V_5, V_6, V_7, V_8, V_9].
\end{equation}

\subsection{Regression-based Stability Constraint Training}
\begin{figure}
    \centering
    \includegraphics[width=\linewidth]{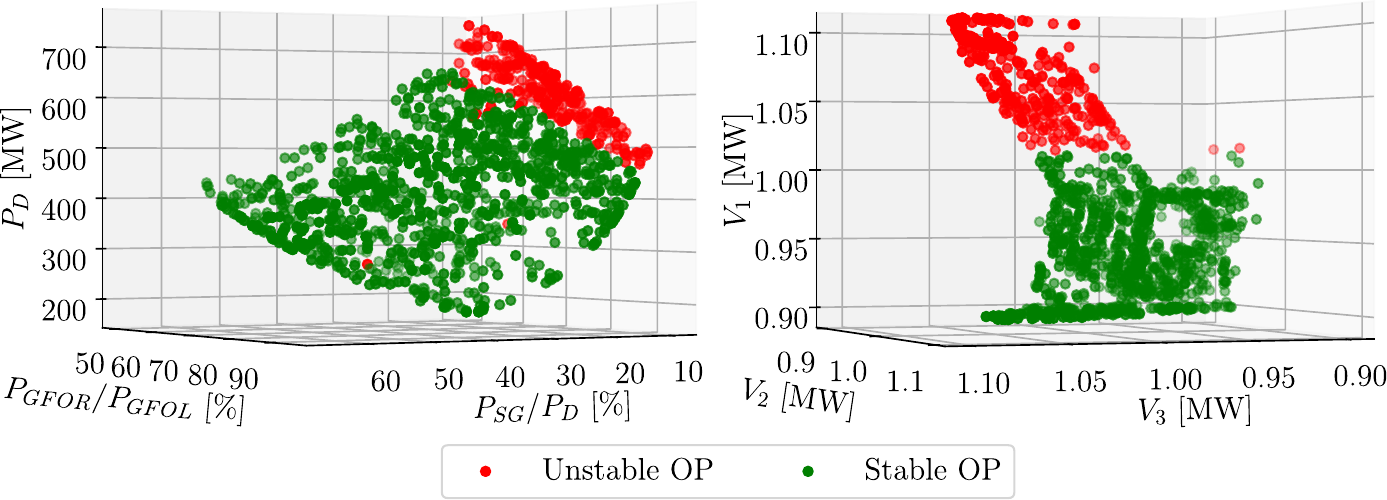}
    \caption{3D scatter plot representation of dataset instances with stability classification.}
    \label{fig:3D_dataset}
\end{figure}

\begin{figure}
    \centering
    \includegraphics[width=\linewidth]{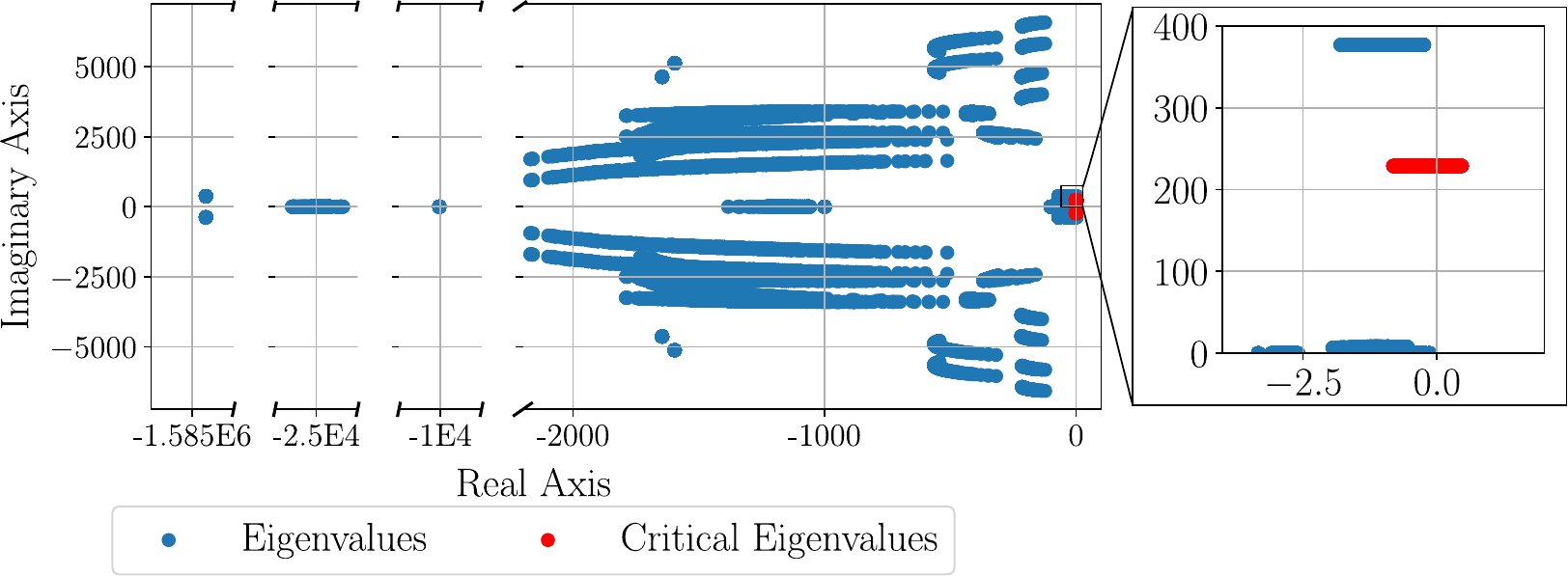}
    \caption{Modal map of the eigenvalues of the operating points in the training dataset.}
    \label{fig:eigenvalues}
\end{figure}

The training dataset is generated using the methodology proposed in~\cite{rossiML2022}, by varying demand and generation within the operating ranges specified in Tables~\ref{tab:load_params}–\ref{tab:gen_parameters}. The stability assessment of each operating point is executed by using the eigenvalues-based analysis provided by the STAMP tool~\cite{arevalo2025matlab}. Figure~\ref{fig:3D_dataset} presents a 3D scatter plot of the dataset instances. On the left, the samples are shown as a function of the total power demand ($P_D$), the share of this demand supplied by the SG, and the distribution of the remaining demand between the GFM and GFL converters. Analysis of these instances indicates that system loading above 400~MW tends to induce instability. Increasing the share of power supplied by the SG mitigates this effect, while, for equal SG participation, a lower contribution from the GFM converter enhances stability. On the right, the training samples are represented as a function of the voltage setpoints of the three generators. This visualization reveals that when the voltage of $\text{GFM}_1$ exceeds 1~p.u., the system becomes unstable. Given this insight into the system’s dynamic behavior, an OPF is formulated in this study with a voltage constraint at bus~1 of $0.9 \leq V_1 \leq 1.0$~p.u., and its solutions are compared with those obtained from the SSSC-OFO and SSSC-OPF formulations, which incorporate the regression-based stability constraint. However, it is important to note that such relationships are relatively easy to identify in this small 9-bus system, whereas in larger networks this approach would be impractical, as these dependencies become much less apparent.
Figure~\ref{fig:eigenvalues} shows the eigenvalues obtained from the small-signal stability analysis of all dataset operating points, with a zoom on the critical eigenvalues that are used for training the regression model. For this group of eigenvalues, the DI is computed as defined in~\eqref{eq:DI} and used as the regression target.
Before training, the data are preprocessed by applying feature selection, retaining as model inputs only uncorrelated variables. Model accuracy is then evaluated using an 80/20 split of the dataset for training and testing, respectively. The resulting $R^2$ score is 0.999 which confirms the suitability of the regression model for this task. Finally, the model is trained using the full dataset. The training process follows the two-step MARS fitting procedure: a forward selection phase and a backward pruning phase~\cite{hastie2009elements}. The pruning step reveals that only two quantities are sufficient to fit a model that accurately computes the DI, namely the voltage magnitudes at buses 1 and 6. The resulting expression is therefore:
\begin{equation}
\begin{split}
    g(\textbf{X}) \;=\;& \;0.0036\,\max\{0,\; 0.9269 - V_6\}\\
    &
    \;-\; 0.0295\,\max\{0,\; 0.9757 - V_1\} \\
    &\;+\; 0.0290\,\max\{0,\; V_1 - 0.9757\}\\
    &
    \;-\; 0.0071\,\max\{0,\; V_6 - 0.9269\}
    \;+\; 0.9991.
\end{split}
\end{equation}

\begin{table}[]
\caption{Regression Accuracy}
\label{tab:mars_accuracy}
\centering
\begin{tabular}{ccc}
\toprule
 & \begin{tabular}[c]{@{}c@{}}Training Data Set\\ (80\% train / 20\% test)\end{tabular} & \begin{tabular}[c]{@{}c@{}}Testing Data Set\\ (OPF solution)\end{tabular} \\ \midrule
$R^2$ & 0.999 & 0.9866\\
\bottomrule
\end{tabular}
\end{table}

\subsection{SSSC-OFO Validation}
%Specific cost function and constraints formulation

The validation procedure is designed to assess the effectiveness of the proposed SSSC-OFO framework in terms of its capability to achieve optimal and dynamically stable solutions, as well as its computational efficiency. To this end, the solutions obtained from the SSSC-OFO are compared with those derived from optimization problems without stability constraints, namely, the conventional OFO and OPF formulations, in order to evaluate the ability of the SSSC-OFO to identify stable solutions when the unconstrained optima are unstable. Furthermore, the performance of the SSSC-OFO is benchmarked against that of the SSSC-OPF to highlight the computational advantages of adopting an OFO-based approach for real-time system operation. Finally, a comparison is carried out against an OPF formulation that includes a voltage constraint at bus~1 ($V_1 \leq 1$~p.u.) to assess whether the regression-based approach provides a more effective means of formulating stability constraints than a simple rule-of-thumb method, which is feasible only for small systems where such relationships are easily identifiable.

\begin{figure}[hpt]
    \centering
    \includegraphics[width=\linewidth]{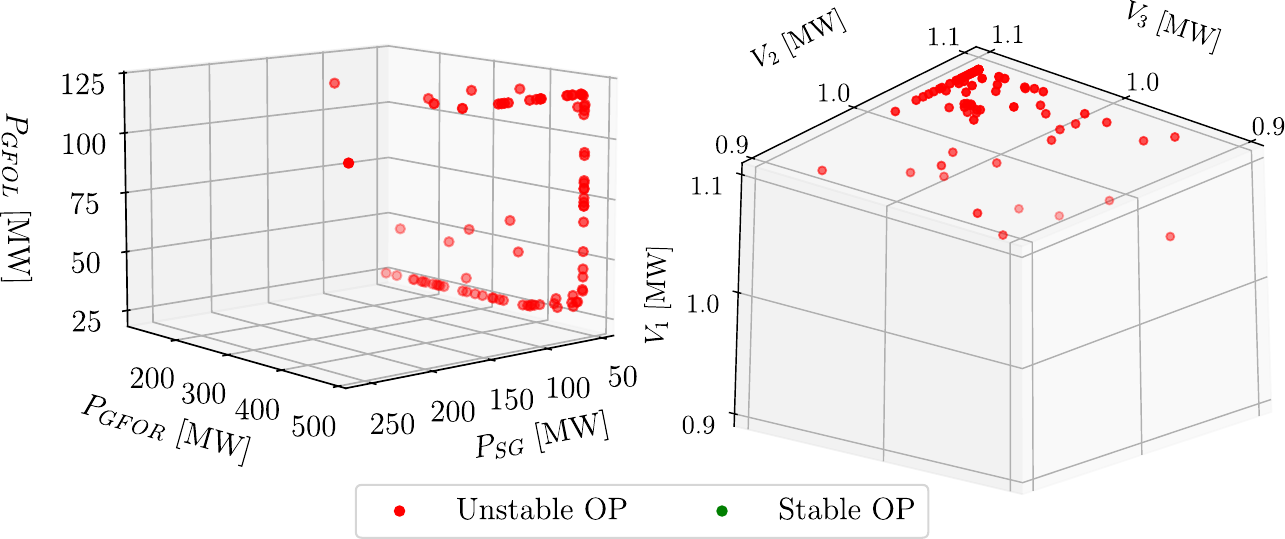}
    \caption{OPF solutions as functions of the generators active power and voltage.}
    \label{fig:3d_test_dataset}
\end{figure}

\begin{figure}[h]
    \centering
    \includegraphics[width=\linewidth]{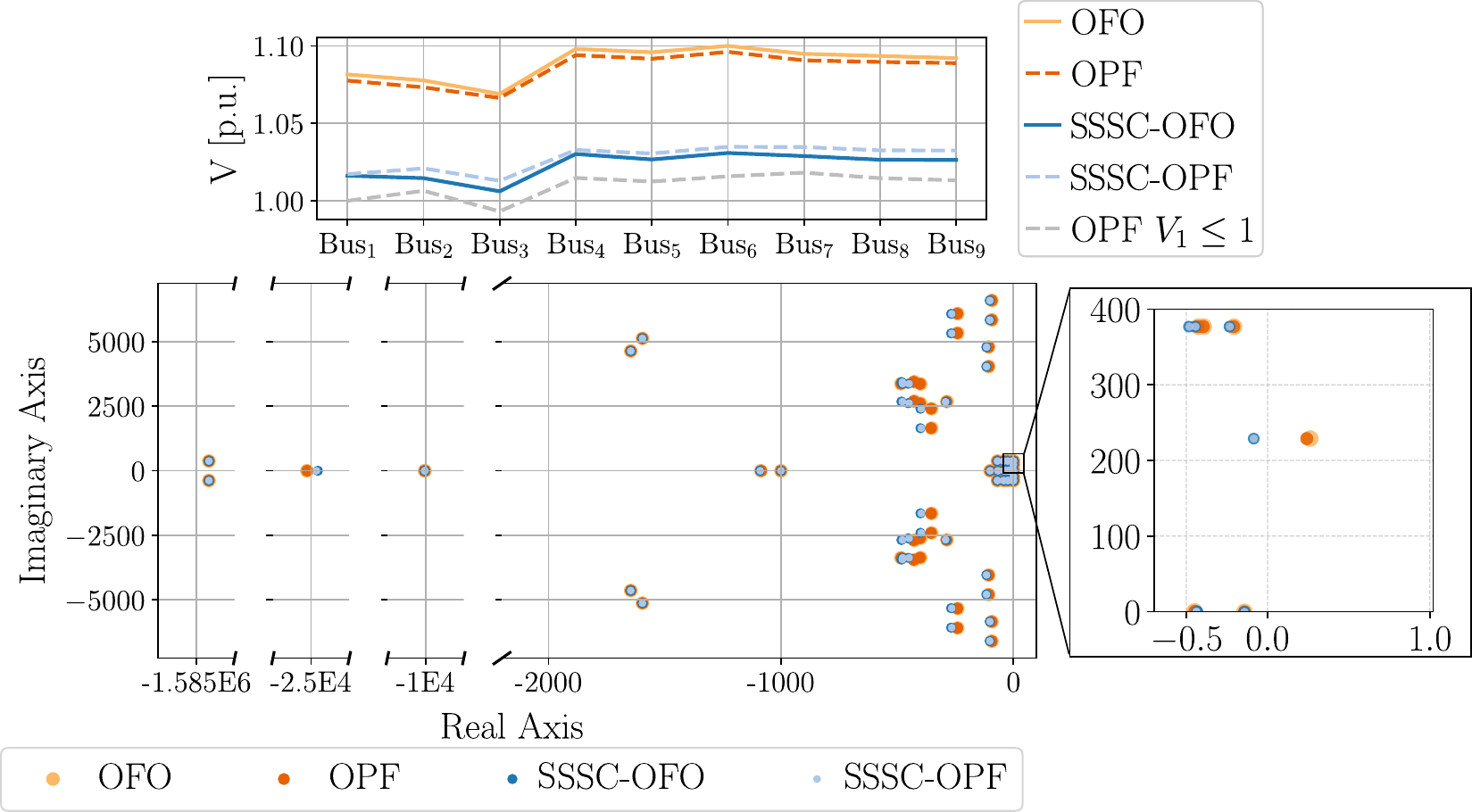}
    \caption{Voltage profiles and modal map of the solutions obtained for the operating point with a low demand level.}
    \label{fig:voltage_modalmap_lowdemand}
\end{figure}

\begin{figure}[h]
    \centering
    \includegraphics[width=\linewidth]{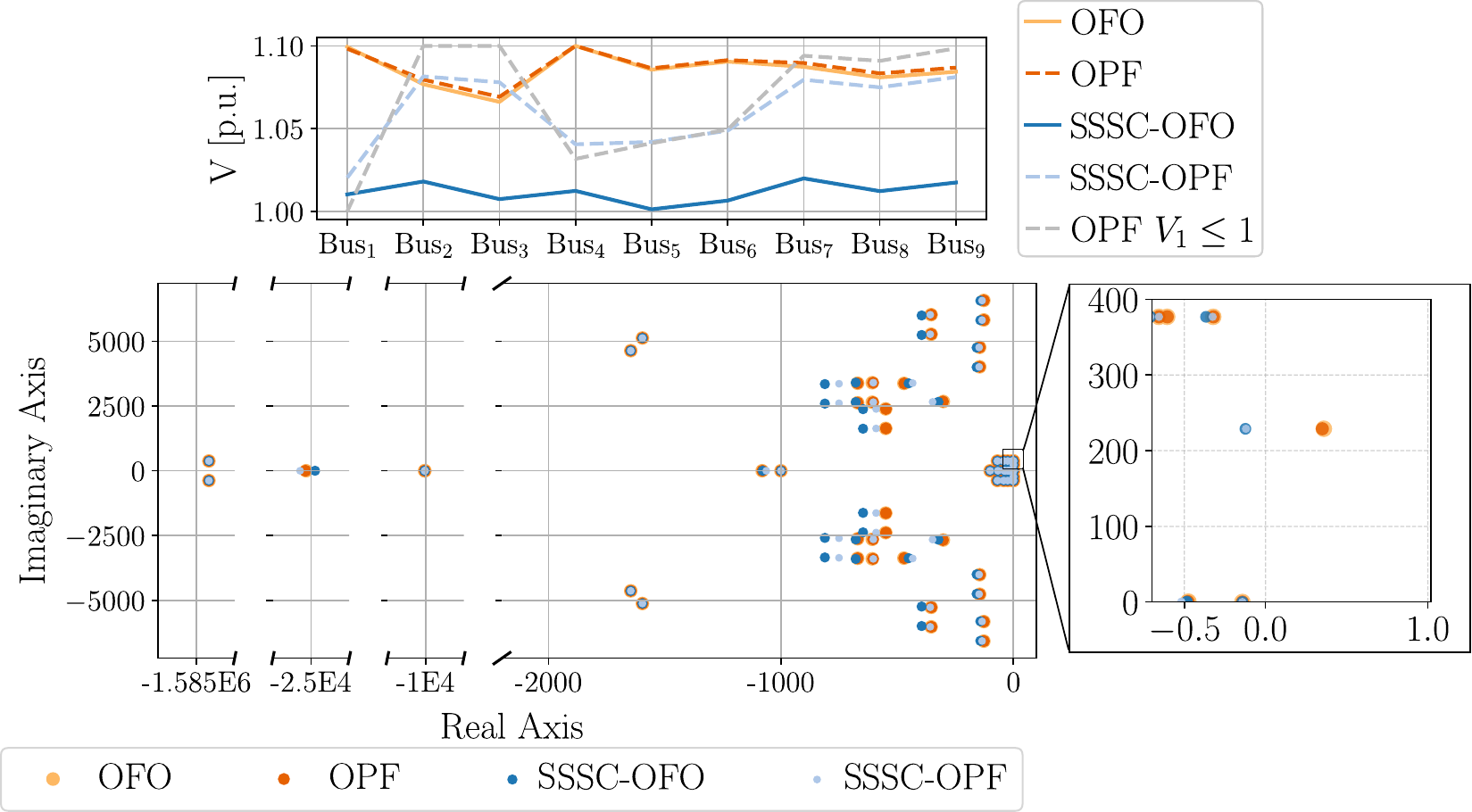}
    \caption{Voltage profiles and modal map of the solutions obtained for the operating point with a medium demand level.}
    \label{fig:voltage_modalmap_mediumdemand}
\end{figure}

\begin{figure}[h]
    \centering
    \includegraphics[width=\linewidth]{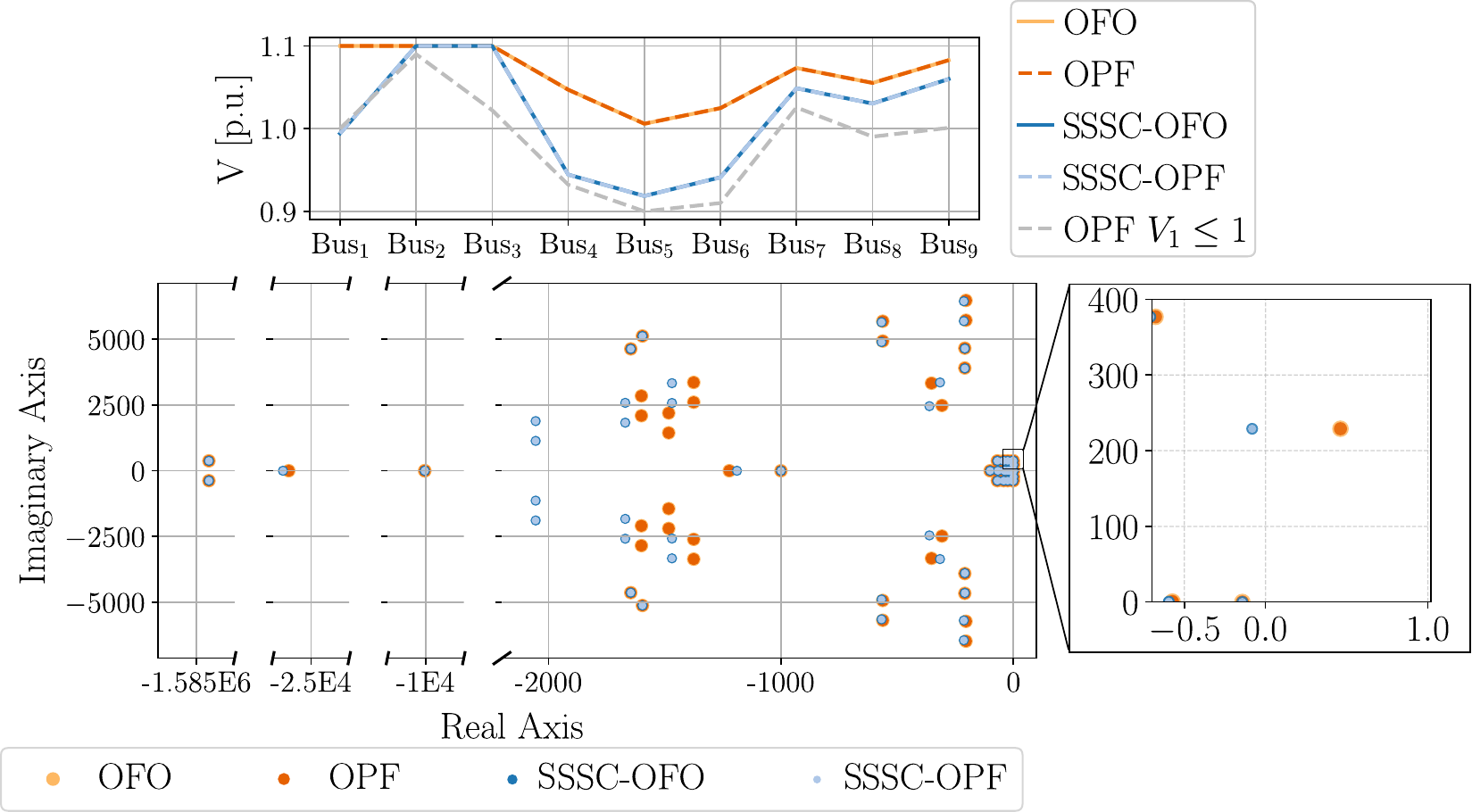}
    \caption{Voltage profiles and modal map of the solutions obtained for the operating point with a high demand level.}
    \label{fig:voltage_modalmap_highdemand}
\end{figure}

Before proceeding with the validation, the system’s stability behaviour is first explored within the region of the operating space associated with the optimal solutions of the problem defined by the objective function in~\eqref{eq:obj_fun_cs}. This preliminary analysis aims to identify the stability characteristics of the operating region relevant to the optimal solutions. To this end, the OPF is solved for 100 operating points uniformly distributed across the system’s operable region, and the small-signal stability of each resulting optimal operating point is evaluated. The obtained results are illustrated in Figure~\ref{fig:3d_test_dataset}. The analysis reveals that, for this specific objective function, the optimal solutions tend to drive the system into unstable operating regions. In order to minimize the total power injected by the generators, the optimization increases the generator voltage levels. Moreover, to further reduce the power contribution from the SG, it is dispatched primarily under high-demand conditions, where its output is required to complement the generation of the other two units, otherwise it is set to its minimum value. Consistent with the behaviour observed in the training dataset, these operating conditions correspond to those that push the system toward instability. A further step in this preliminary analysis is to evaluate the regression accuracy in estimating the DI of the OPF solutions. As reported in Table~\ref{tab:mars_accuracy}, the obtained $R^2$ value of 0.9866 confirms the suitability of the regression model for use as a stability constraint, enabling the subsequent validation analysis.
% Please add the following required packages to your document preamble:
% \usepackage{multirow}
\begin{table*}[t]
\caption{Solutions description summary.}
\label{tab:sol_summary}
\begin{tabular}{ccccccccccc}
\hline
Case &  & \multicolumn{1}{c}{\begin{tabular}[c]{@{}c@{}}$P_{GFM_1}$\\ MW\end{tabular}} & \multicolumn{1}{c}{\begin{tabular}[c]{@{}c@{}}$P_{SG_2}$ \\ MW\end{tabular}} & \multicolumn{1}{c}{\begin{tabular}[c]{@{}c@{}}$P_{GFL_3}$\\ MW\end{tabular}} & \multicolumn{1}{c}{\begin{tabular}[c]{@{}c@{}}$Q_{GFM_1}$\\ Mvar\end{tabular}} & \multicolumn{1}{c}{\begin{tabular}[c]{@{}c@{}}$Q_{SG_2}$\\ Mvar\end{tabular}} & \multicolumn{1}{c}{\begin{tabular}[c]{@{}c@{}}$Q_{GFL_3}$ \\ Mvar\end{tabular}} & \multicolumn{1}{c}{$f(\mathbf{y},\mathbf{u})$} & \multicolumn{1}{c}{\begin{tabular}[c]{@{}c@{}}DI\\ by MARS\end{tabular}} & \multicolumn{1}{c}{\begin{tabular}[c]{@{}c@{}}DI\\ Exact\end{tabular}} \\ \hline
\multirow{5}{*}{\begin{tabular}[c]{@{}c@{}}Low\\ Demand Case\\ ($P_D = 207.23$ MW)\end{tabular}} & OPF & 128.24 & 54 & 25 & -26.71 & -29.09 & -40.77 & 0.5027 & 1.0008 & 1.0011 \\
 & OFO & 128.23 & 54 & 25 & -26.96 & -28.78 & -42.04 & 0.5027 & 1.0009 & 1.0011 \\
 & \textbf{SSSC-OFO} & \textbf{128.37} & \textbf{54} & \textbf{25} & \textbf{-20.12} & \textbf{-22.26} & \textbf{-34.54} & \textbf{0.5029} & \textbf{0.9995} & \textbf{0.9996} \\
 & SSSC-OPF & 128.36 & 54 & 25 & -23.11 & -21.6 & -33.6 & 0.5028 & 0.9995 & 0.9996 \\
 & \multicolumn{1}{l}{OPF $V_1 \leq$1} & 128.41 & 54 & 25 & -21 & -17.96 & -33.79 & 0.5029 & 0.9991 & 0.9992 \\ \hline
\multirow{5}{*}{\begin{tabular}[c]{@{}c@{}}Medium\\ Demand Case\\ ($P_D = 318.55$ MW)\end{tabular}} & OPF & 239.54 & 54 & 25 & 10.45 & -16.64 & -32.08 & 0.6589 & 1.0014 & 1.0015 \\
 & OFO & 239.55 & 54 & 25 & 12.54 & -17.38 & -32.95 & 0.6589 & 1.0015 & 1.0016 \\
 & \textbf{SSSC-OFO} & \textbf{240.29} & \textbf{54} & \textbf{25} & \textbf{12.61} & \textbf{-2.19} & \textbf{-17.08} & \textbf{0.6603} & \textbf{0.9995} & \textbf{0.9995} \\
 & SSSC-OPF & 240.30 & 54 & 25 & 11.20 &	-9.48 &	-9.33 &	0.6603		&0.9995 &	0.9995 \\
 & \multicolumn{1}{l}{OPF $V_1 \leq$1} & 223.21 & 54 & 42.08 & -41.07 & 11.23 & 3.26 & 0.7034 & 0.9989 & 0.9989 \\ \hline
\multirow{5}{*}{\begin{tabular}[c]{@{}c@{}}High\\ Demand Case\\ ($P_D = 661.41$ MW)\end{tabular}} & OPF & 486.40 & 56.32 & 118.70 & 161.28 & 47.73 & 35.90 & 0.2239 & 1.0020 & 1.0020 \\
 & OFO & 486.40 & 56.27 & 118.75 & 161.29 & 47.73 & 35.91 & 0.2239 & 1.0020 & 1.0020 \\
 & \textbf{SSSC-OFO} & \textbf{486.40} & \textbf{62.56} & \textbf{118.75} & \textbf{160.21} & \textbf{91.22} & \textbf{78.82} & \textbf{0.2342} & \textbf{0.9995} & \textbf{0.9996} \\
 & SSSC-OPF & 486.40 & 62.61 & 118.69 & 160.30 & 91.32 & 78.70 & 0.2342 & 0.9995 & 0.9996 \\
 & \multicolumn{1}{l}{OPF $V_1 \leq$1} & 486.40 & 63.78 & 117.34 & 192.57 & 113.36 & 40.94 & 0.2342 & 0.9998 & 1.0000 \\ \hline
\end{tabular}
\end{table*}
Given that all the optimal solutions of this problem are small-signal unstable, the performance of the SSSC-OFO is examined in greater detail by selecting three operating points representative of different demand levels. Table~\ref{tab:sol_summary} summarizes for each case, the obtained active and reactive power dispatch of the generators, the corresponding objective function value $\phi(\mathbf{u}, \mathbf{y})$, and the DI values computed both by the regression model and by the exact eigenvalue analysis of the linearized state-space model of the system. Figures~\ref{fig:voltage_modalmap_lowdemand}–\ref{fig:voltage_modalmap_highdemand} present detailed results for the three operating cases, including the modal maps and voltage profiles. It can be observed that, for all demand levels, the OFO and OPF solutions coincide, thereby confirming the consistency and correctness of the OFO formulation. Similarly, the SSSC-OFO and SSSC-OPF solutions exhibit almost identical results, with matching active power dispatch among generators, very similar voltage profiles, and stable operating points. Finally, the solutions of the OPF with the voltage constraint at bus~1 indicate that this formulation does not always yield an optimal solution (as in the medium-demand case) or a stable one (as in the high-demand case).\\
Table~\ref{tab:iterations} reports the number of iterations required by each model to reach convergence under different demand levels. The results show that the proposed SSSC-OFO achieves convergence with a significantly lower number of iterations compared to the unconstrained OFO, particularly at medium and high demand levels. This demonstrates the effectiveness of incorporating the regression-based small-signal stability constraint, which guides the optimization toward feasible and dynamically stable regions, improving numerical convergence. In contrast, the conventional OFO requires many more iterations, especially in the low- and medium-demand cases, where instability or constraint violations hinder convergence. The SSSC-OPF exhibits comparable or even faster convergence than SSSC-OFO, as expected for an offline solver, while the standard OPF and voltage-constrained OPF show inconsistent iteration counts across demand levels.%, reflecting the absence of dynamic stability awareness. 
\begin{table}[h!]
\centering
\caption{Number of iterations for each model and demand level}
\label{tab:iterations}
\begin{tabular}{lccc}
\toprule
\multirow{2}{*}{Model / Case} & \multicolumn{3}{c}{Demand Level} \\ \cmidrule(lr){2-4}
 & Low & Medium & High \\ 
\midrule
OPF & 106 & 15 & 21 \\
OFO & 352 & 63 & 11 \\
SSSC-OFO & 73 & 47 & 4 \\
SSSC-OPF & 11 & 51 & 23 \\
OPF ($V_1 \leq 1$) & 16 & 76 & 18 \\
\bottomrule
\end{tabular}
\end{table}

Overall, the SSSC-OFO provides a favorable trade-off between stability enforcement and computational efficiency, making it suitable for online operation.
Appendix~\ref{appendix:opt_params_tuning} reports the parameters $\gamma$ and $\mathcal{G}$, which were tuned to enhance the performance of each model and ensure a fair comparison among the different solutions.

%\subsection{Industrial case}

%The proposed methodology is tested on a modified IEEE 118-bus system.

%\subsubsection{Stability Regression Analysis}

%\subsubsection{Real Time Operation}

% chapter 2 pag 37 model ieee 9 bus system

\section{Conclusions and future work} \label{sec:concl}
This paper presented a SSSC-OFO framework enabling real-time optimal operation of power systems while explicitly accounting for dynamic stability margins. The proposed methodology integrates a data-driven regression-based formulation of the small-signal stability constraint into an OFO structure, allowing fast evaluation and enforcement of stability conditions during operation. Simulation results on a modified IEEE 9-bus system demonstrated that the SSSC-OFO reliably guides the system toward stable and near-optimal operating points, succeeding in ensuring stability in cases where optimization without dynamic restrictions would lead to unstable operating conditions, while achieving comparable or reduced computational effort.

Future work will focus on extending the proposed methodology to larger and more realistic power systems to further evaluate scalability and performance. In addition, efforts will be directed toward adapting the formulation for application to hybrid AC/DC networks, particularly in the presence of HVDC links. %Additional efforts will include real-time validation using hardware-in-the-loop (HIL) testing to evaluate the controller’s behavior under practical implementation constraints.

%Intro reescrita: falta completar técnicas Data-Driven y definir cuales son los puntos novedosos.
%Propuesta: falta MARS, el OFO ya lo he explicado
%Análisis simulación: he descrito el caso para red de 9 buses. Hay que discutir como presentar los resultados.
%Conclusiones: pendiente

%Referencia a la red de 9 buses
%Cómo operan los generadores? (importa para la regresión)
%Cómo se modelan las cargas?
%Las cargas siguen definidas con la reactiva a 0.0?
%Cómo presentar los resultados?
%	MARS -> ecuación obtenida (?)
%	SOFO -> Simulación dinámica y Eigenvalues en el estado final (?)

\appendices
\section{Optimal Power Flow and Online Feedback Optimization Parameters Tuning}
\label{appendix:opt_params_tuning}
\begin{table}[h!]
\caption{Parameters of low demand case.}
\label{tab:low_demand_parameters}
\centering
\begin{tabular}{lll}
\toprule
Model & $\gamma$ & $\mathbf{G}$ \\ 
\midrule
OPF & 1000 & -- \\
SSSC-OPF & 1000 & -- \\
OFO & 100 & $\mathrm{diag}(1,\, 1,\, 0.1,\, 0.1,\, 0.1)$ \\
SSSC-OFO & 100 & $\mathrm{diag}(1,\, 1,\, 0.1,\, 0.1,\, 0.1)$ \\
OPF $V_1 \leq 1$ & 100 & -- \\ 
\bottomrule
\end{tabular}
\end{table}

\begin{table}[htp!]
\caption{Parameters of the medium and high demand cases}
\label{tab:med_high_demand_parameters}
\centering
\begin{tabular}{cccc}
\toprule
\multirow{2}{*}{Model} & \multicolumn{2}{c}{$\gamma$} & \multirow{2}{*}{$\mathbf{G}$} \\ 
\cmidrule(lr){2-3}
 & Medium & High &  \\ 
\midrule
OPF & \multicolumn{2}{c}{100} & -- \\
SSSC-OPF & 10 & 100 & -- \\
OFO & \multicolumn{2}{c}{100} & $\mathrm{diag}(1,\, 1,\, 0.2,\, 0.2,\, 0.2)$  \\
SSSC-OFO & \multicolumn{2}{c}{100} & $\mathrm{diag}(1,\, 1,\, 0.2,\, 0.2,\, 0.2)$ \\
OPF $V_1 \leq 1$ & \multicolumn{2}{c}{100} & -- \\ 
\bottomrule
\end{tabular}
\end{table}

% Appendix one text goes here.

% \section{}
% Appendix two text goes here.

%\section*{Acknowledgment}
%This work was supported by the Project TED2021-130351B-C21 (HP2C-DT), funded by MICIU/AEI/10.13039/501100011033 and by the European Union NextGenerationEU/PRTR.
%The authors would like to thank...

% \ifCLASSOPTIONcaptionsoff
%   \newpage
% \fi

\bibliographystyle{ieeetr}
\bibliography{bibtex/bib}

% \begin{IEEEbiography}{Michael Shell}
% Biography text here.
% \end{IEEEbiography}

% \begin{IEEEbiographynophoto}{John Doe}
% Biography text here.
% \end{IEEEbiographynophoto}

% \begin{IEEEbiographynophoto}{Jane Doe}
% Biography text here.
% \end{IEEEbiographynophoto}

\end{document}